\documentclass[titlepage,12pt]{utarticle}
\usepackage{amssymb,amscd}
\input epsf.tex
\numberwithin{equation}{section}

\newcommand{\ket}[1]{\lvert #1\rangle}
\newcommand{\kket}[1]{\lvert #1\rangle\rangle}

\newcommand{\one}{{\bf 1}}

\newcount\figno
\def\fig#1#2#3{
\par\begingroup\parindent=0pt\leftskip=1cm\rightskip=1cm\parindent=0pt
\baselineskip=11pt
\global\advance\figno by 1
\epsfxsize=#3
\centerline{\epsfbox{#2}}
\vskip 12pt
{\bf Figure \the\figno:} #1\par
\endgroup\par
}
\def\figlabel#1{\xdef#1{\the\figno
\mbox{ }}}
\def\encadremath#1{\vbox{\hrule\hbox{\vrule\kern8pt\vbox{\kern8pt
\hbox{$\displaystyle #1$}\kern8pt}
\kern8pt\vrule}\hrule}}

\begin{document}
\preprint{
  CERN--TH/2001-290\\
  {\tt hep-th/0110219}\\
}
\title{ On Orientifolds of WZW Models \\{~}\\ and their Relation to Geometry
}
\author{
  Ilka Brunner
    \oneaddress{
     Theory Division\\
     CERN\\
     CH-1211 Geneva 23\\
     Switzerland\\
     {~}\\ 
     \email{ilka.brunner@cern.ch}
    }
}

\Abstract{
We investigate D-branes in orientifolds of WZW models. A connection
between the conformal field theory approach to orientifolds and the
target space motivated analysis is established. 
In particular, we associate previously constructed crosscap states to
involutions of the group manifold and their fixed point sets.
Whereas our analysis of D-branes in orientifolds of general WZW
models is restricted to special D0-branes, we investigate all
symmetry preserving branes of SU(2)-orientifolds in detail.
For that case, the location of the orientifold fixed point set is independently
determined by scattering localized graviton wave packets.\\[3cm]
CERN--TH/2001-290\\[0.2cm]
October 2001
}
\date{}

\maketitle

\section{Introduction}
In the last two years considerable progress has been achieved in
the understanding of D-branes on group manifolds
\cite{AlSc,ARSI, ARSII,FFFS,MMSI,Stan,KlSe}, in particular on
the geometrical interpretation of  conformal field
theory results. In this paper, we start a similar program
for orientifolds of group manifolds.

In conformal field theory, D-branes are characterized by boundary
conditions for the elements of the chiral algebra. For maximally
symmetric gluing conditions $J(z) = \bar J(z)$ on the upper half plane
an exact conformal field theory description in terms of boundary
states had been given starting with the work of Cardy.
More recently, it was shown how the geometry of the associated
D-branes is encoded in the gluing conditions. In particular,
the gluing conditions constrain the end points of open strings
to  conjugacy classes of the group manifold \cite{AlSc}. 
\medskip

In the case of orientifolds, the underlying theory is modded
out by an involution involving world sheet parity and possible
additional actions on the target manifold. The target space action
can in general have fixed points. In flat space, we associate
to the fixed points of such involutions ``orientifold planes''
in the target geometry. In this way, orientifolds  have a geometrical
meaning in terms of target data. 

In conformal field theory, orientifolds are described by crosscap
states. The construction of crosscaps
on group manifolds and in general rational conformal
field theory has been investigated in \cite{PSSI, PSSII, PSSIII,
HSSI, HSSII, FHSSW, horava}. 
In this paper we  shall match the conformal field
theory data of the orientifold encoded in a choice of
consistent crosscap state to the geometric description
in terms of a target space involution.

The first part of the paper is devoted to the case of
an arbitrary WZW model.
We discuss suitable target involutions that preserve all symmetries,
and review
some essential points of the work of 
\cite{PSSI, PSSII, PSSIII, HSSI, HSSII, FHSSW, FFFSII, FuSc, SaSt, BiSt} 
on crosscap states. We then map crosscap states 
to suitable target involutions, thus connecting the world sheet and target
space motivated analysis. The tool is the interpretation of
special M\"obius strip amplitudes.

In the second part, we investigate orientifolds of $SU(2)$ in greater
detail. For $SU(2)$ the conjugacy classes
are generically two-spheres in $S^3\sim SU(2)$ and two additional conjugacy
classes are points. These conjugacy classes can be wrapped
by symmetry preserving D-branes.
For $SU(2)$ there
are two different target involutions which can be divided
out while preserving the $SU(2)$ current algebra. Both of
these have fixed point sets, to which orientifold loci
(which are not planes) can be associated. 

The reflections of $S^3$  act naturally on the open string sector
of the theory. Geometrically, one can determine how
conjugacy classes transform under the involution. 
If the conjugacy class is mapped to itself, the
reflection induces an involution on the algebra of functions
on the conjugacy class, which is the algebra of spherical
harmonics. From a conformal field theory point of view,
this induced involution corresponds to the action of $\Omega$
in the open string sector with boundary conditions determined
by the conjugacy class. We compare the geometrically induced
involution to the action of $\Omega$ as determined by the
crosscap and boundary state for all boundary conditions.


The location of the orientifold can be determined independently
by scattering localized graviton wave packets from it.
We perform this computation in the case of $SU(2)$, confirming
the results of the earlier sections.

\section{General considerations}
\subsection{
Fixed point sets and  gluing conditions}
The symmetries of a WZW model consist of a left- and a right-moving
current algebra. In terms of group elements, the currents can be
written as
\begin{equation} 
J= k \,g^{-1} \partial g, \quad \quad \bar{J} = -k \, \bar\partial g \, g^{-1}.
\end{equation}
For a symmetry preserving orientifold induced by an involution
$\Omega$,  the left
and right moving current algebras must be interchanged under
the  orientifold action.
In the flat space case one is used to thinking of orientifolds
in terms of orientifold planes in the target geometry. 
The aim of the following sections is to find 
a similar geometrical interpretation 
in terms of the target space data also when
the target space is a group manifold.
Stated differently, one requires an involution on the target, such that the
induced action on the currents is to exchange the left and right
moving current algebra. The fixed point set
is then interpreted geometrically as the location of the orientifold.

An example for such an involution is
\cite{BiSt, FGP,John}:
\begin{equation}
\omega_+:  g(z,\bar z) \to  g^{-1} (\bar z, z) .
\end{equation}
This involution exists generically for every WZW model.
It maps conjugacy classes to conjugacy classes. The precise
form of the fixed point set depends on the particular group
under consideration. The identity $\one$ of the group is 
always part of the fixed point set, but the involution
might leave other conjugacy classes fixed.

In addition, there can exist other involutions  which can be
divided out. If the group manifold has a non-trivial center ${\cal Z}$, then
the following involution can also be considered:
\begin{equation}\label{involution}
\omega_z:  g(z,\bar z) \to  z g^{-1} (\bar z, z) \quad z \in {\cal Z}
\end{equation}
In order to map $J$ and $\bar J$ on each other (the case we want to consider
in this paper), $z$ has to square to the identity,
$z^2 = \one$. The fixed point set consists of all group elements
squaring to $z$. In general, it can have
several disconnected components consisting of various Grassmannians.

\subsection{Crosscap states for WZW models}

Starting from a closed string vacuum whose closed string
spectrum is encoded in the charge conjugation modular invariant,
open descendants containing unoriented strings
have been constructed in 
\cite{PSSI, PSSII, PSSIII, HSSI, HSSII, FHSSW, SaSt, FFFSII}.
These provide candidates for a conformal field theory description
of orientifolds, to which we later want to associate the
involutions discussed above.
We will therefore briefly review the relevant
formulas from those papers. 
The construction of orientifolds involves unorientable
worldsheets. The basic ingredient is the crosscap $\BR \BP_2$, the simplest
non-orientable surface. The introduction of the crosscap breaks
the ${\cal A} \times {\cal \bar A}$ symmetry of the model, where
${\cal A}$ (${\cal \bar A}$) denote the left (right) moving current
algebra. Here, we are considering crosscaps that leave one
copy of the current algebra unbroken. The condition for this is
that
\begin{equation}\label{cblock}
\left( J_n - (-1)^n \bar{J}_{-n} \right) \ket{C,j} = 0 
\end{equation}
The $\ket{C,j}$ are crosscap Ishibashi states  
normalized such that 
\begin{equation}\label{norm}
<C,j| e^{-\frac{\tilde t}{2} 
(L_0+\tilde L_0)}| C,l> = \delta_{j,l} \ \chi_j (i\tilde t)
\end{equation}
Uncapitalized letters  $j,l, \dots$ label the representations of the chiral
algebra, in the case at hand the irreducible representations
of the current algebra ${\cal A}$.
Crosscap states are then linear combinations of the crosscap
Ishibashi states leading to consistent Klein bottle and
M\"obius strip amplitudes. A formula for consistent crosscaps
in the context of RCFT has been given by \cite{SaSt, PSSIII}
\begin{equation}\label{CI}
\ket{C_{gen}} = \sum_j \frac{P_{j0}}{\sqrt{S_{j0}}} \ \ket{C,j},
\end{equation} 
where
\begin{equation}\label{P}
P = T^{\frac{1}{2}}  \ S  \ T^2 \ S \ T^{\frac{1}{2}}
\end{equation}
The transverse channel of the Klein bottle describes the propagation
of closed strings on a tube terminating on two crosscaps. This
amplitude follows immediately from (\ref{norm}) and (\ref{CI}). 
Using an $S$ modular transformation, one obtains the direct
channel of the Klein bottle:
\begin{equation}
K = \sum_i \ Y_{00}^i \ \chi_i,
\end{equation}
where $Y_{00}^i$ is the Frobenius Schur indicator \cite{HSSII, BantII}
introduced for RCFT in \cite{BantI}. 
It is $1$ when $i$
is a real representation, $-1$ if $i$ is pseudo-real 
and $0$ if it is complex \cite{BantI}. $Y_{00}^i$ is a special component of
the following integer valued \cite{PSSIII, gann} tensor:
\begin{equation}
Y_{ij}^k = \sum_m \frac{S_{im} \ P_{jm} \ P_{km}^\dagger}{S_{0m}}.
\end{equation}
This crosscap is consistent with all Cardy boundary states. 
The amplitudes corresponding to the M\"obius strip are in the tree
channel given by cylinders capped off by a crosscap at one
end and by a boundary state at the other end. The 
transformation of these amplitudes to the open string
sector results in a one-loop amplitude with an $\Omega$
insertion in the open string sector. The matrix mediating between
the two channels is the matrix $P$ of equation (\ref{P}).
\medskip

We have seen that in the case that
the group has a non-trivial center, in particular if the center
contains a $\BZ_2$ subgroup, there are additional possibilities
for orientifolds. In conformal field theory, the existence
of a non-trivial abelian center, for example  $\BZ_N$ for $SU(N)$,
manifests itself in the existence of
simple currents. These are special primaries whose fusion with
any other field yields exactly one field. The abelian group generated
by these currents is isomorphic to the center of the group.
From this point of view, it has to be expected that
simple currents enter the discussion of orientifolds. Indeed,
additional consistent crosscap states in models with
non-trivial simple current groups where discovered in \cite{HSSII, PSSII}.
The explicit expression for a crosscap involving a simple
current $L$  is \cite{HSSII}
\begin{equation}
\ket{C_L}= \sum_j \ \frac{P_{jL}}{\sqrt{S_{j0}}} \ \ket{C,j},
\end{equation}
leading to the modified Klein bottle
\begin{equation}
K_L = \sum_i Y_{LL}^i \ \chi_i = 
\sum_i e^{2\pi i Q_L(i)} \ \ Y_{00}^i \  \chi_i.
\end{equation}
Here, $Q_L(i)$ is the monodromy charge of the field $i$ with
respect to the current $L$:
\begin{equation}
Q_L(i) = h_L + h_i - h_{L \times i} \quad  {\rm mod} \ 1
\end{equation}

\subsection{Matching Crosscap states and fixed point sets}

A priori, it seems
suggestive that the crosscap $\ket{C_{gen}}$ in equation (\ref{CI}), which
exists in any WZW model should provide the world sheet description of
the target involution $g \to g^{-1}$, which exists for all group
manifolds. The crosscap $\ket{C_L}$ should correspond to the
involution $g \to z g^{-1}$, as can be expected from the
connection between simple current groups and non-trivial centers
of groups. 

This can be confirmed by the following argument: Consider the brane
located at the identity. For any group manifold, the identity represents
a conjugacy class consisting of one point. The D-brane sitting
at that point is described by the Cardy state labeled by the
vacuum character
\begin{equation}
\ket{0} = \sum_i \ \sqrt{S_{0i}} \ \kket{i},
\end{equation}
where $\kket{i}$ is the Ishibashi state built on the primary $i$.
The open string channel of the cylinder with those boundary
conditions contains only the character $\chi_0$. 

The involution $g \to g^{-1}$ leaves the identity fixed, and
therefore, the image of the brane $\ket{0}$ is $\ket{0}$.
The M\"obius amplitude can be interpreted as an amplitude
between the brane and its image. In the present situation, it 
can be concluded that the character $\chi_0$ is also the only
one appearing in the M\"obius amplitude. Using the matrix $P$,
one obtains the transverse amplitude
\begin{equation}
M_0 = \chi_0 \longleftrightarrow \sum_j P_{0j} \chi_{j}
= \sum_j \sqrt{S_{0j}} \ \Gamma_j \chi_{j}.
\end{equation}
In the last step, the amplitude in the transverse channel was
expressed as an overlap of the boundary state $\ket{0}$ and
a crosscap $\ket{C} = \sum \Gamma_j \ket{C,j}$. It can be read off
that the coefficient of the crosscap $\ket{C}$ are given
by
\begin{equation}
\Gamma_j = \frac{P_{0j}}{\sqrt{S_{0j}}}.
\end{equation}
This supports that the crosscap $\ket{C_{gen}}$ given in the previous
section corresponds to the fixed point set of $g \to g^{-1}$.
Of course, there is at this point of the discussion a choice of an overall
sign, and also the crosscap $-\ket{C_{gen}}$ would be located
at the same fixed point set. The sign reverses the projection on
$\Omega$-invariant states in the open string sector and would
be determined in a full string theory by tadpole cancellation
conditions.

The involution $g \to z g^{-1}$ maps the identity to the group
element $z$. The D-brane located at $z$ is represented by the
Cardy state which carries the representation label of the simple current $L$
corresponding to $z$:
\begin{equation}
\ket{L} = \sum_j e^{2\pi i Q_L(j)} \ \sqrt{S_{0j}} \ \ket{j}
\end{equation}
To compare to Cardy's standard formula, note that
$S_{Lj} = \exp(2\pi i Q_L(j) ) S_{0j}$ \cite{ScYa}.
The open string spectrum with boundary conditions $\ket{0}$ and
its image $\ket{L}$ consists of the single character $\chi_L$. As before,
it can be concluded that the only character appearing in
the one loop channel of the M\"obius amplitude is $\chi_L$
\begin{equation}
M^L_0 =\chi_L \longleftrightarrow \sum P_{Lj} \chi_{j}
= \sum_j \sqrt{S_{0j}} \ \Gamma_j \chi_{j}. 
\end{equation}
The factorization in the transverse channel gives 
\begin{equation}
\Gamma_j = \frac{P_{Lj}}{\sqrt{S_{0j}}}.
\end{equation}
For consistency, since $z$ and $\one$ are images of {\it each other},
the M\"obius strip with boundary condition $\ket{L}$ equals the one with
boundary condition $\ket{0}$ and accordingly it must be possible to factorize
the amplitude in the transverse channel also into the crosscap state
and the boundary state $\ket{L}$. This is indeed possible,
since $P_{Lj}=0$ whenever $Q_L(j) \in \BZ + 1/2$ (and L is (half-)
integer spin). 

The conclusion we can reach is therefore that the involution
$g \to z g^{-1}$ corresponds to the crosscap state $\ket{C_L}$,
whereas $\ket{C_{gen}}$ corresponds to $g \to g^{-1}$.

Turning the chain of arguments around, the above discussion could lead
to an alternative derivation of the coefficients of the crosscap 
state: One starts
with an involution on the group manifold and determines the spectrum
of open string states and the action of $\Omega$
for a particular boundary condition. Here, we have chosen the boundary
condition $\ket{0}$. One then transforms to the transverse channel
and factorizes the result into a boundary and a crosscap state.
The coefficients for the crosscap state can be read off from
the amplitude. In principle, one can perform this type of analysis
for any boundary condition, and the analysis of those
leads to important consistency checks. In the case of
$SU(2)$, which we consider in the next section, we will in fact
study all boundary conditions, leading to a detailed match of
all amplitudes with geometrical expectations.

\section{Example: $SU(2)$}

\subsection{Involutions and crosscaps}

Let us apply these considerations to the case of $SU(2)_k$, where
throughout the rest of this paper it is assumed that $k$
is even. See \cite{BiSt, John, FGP} for earlier discussions of $SU(2)$
orientifolds in the context of five-branes and \cite{horava}
for an early  discussion of crosscap states under consideration
in this section.

The center of $SU(2)$ consists of the two elements $\{ \one, -\one \}$.
Accordingly, there  are two possible involutions (\ref{involution})
which can be used for orientifolds.
The fixed point set of the inversion $g \to g^{-1}$ consists
of $\pm \one$, whereas the fixed point set of $g \to - g^{-1}$
consists of the conjugacy class of the group element
\begin{equation}
g_{eq} = \left( \begin{array}{cc} e^{\frac{\pi i}{2}} & 0 \\ 
                                 0& e^{-\frac{\pi i}{2}} 
                 \end{array} \right),
\end{equation}
which is
the conjugacy class located at the equator of $S^3$.

Turning to the CFT point of view, recall that the characters
of $SU(2)_k$ are labeled by $j= 0, 1/2, 1, \dots k/2$. 
Accordingly, there are $k+1$ Cardy states $\ket{J}$ wrapping
the $k+1$ integral conjugacy classes that exist at
finite $k$. More precisely, the Cardy state $\ket{J}$ wraps
the conjugacy class represented by
\begin{equation}\label{gj}
g_J = \left( \begin{array}{cc} e^{\frac{2\pi i J}{k}} & 0 \\
                              0 & e^{-\frac{2\pi i J}{k}}
             \end{array}
      \right).
\end{equation}

The non-vanishing entries of the matrix $P$ for $SU(2)_k$ ($k$ even) are 
given by
\begin{equation}
P_{jl}= \frac{2}{\sqrt{k+2}} \sin\frac{\pi (2j+1)(2l+1)}{2(k+2)}
 \quad for \quad j+l \in \BZ 
\end{equation}
There is a simple current in the theory which carries the
representation  $k/2$ and maps the primaries $j$ to
$k/2 -j$. The simple current group $\{ 0, k/2 \}$ is
therefore $\BZ_2$ and  isomorphic to the center of the group.
According to the above discussion, the crosscap state located
at the equator of the group manifold is
\begin{equation}
\ket{C_{eq}} = \left(\frac{2}{k+2} \right)^{\frac{1}{4}}
\sum_{j=0}^{k/2} (-1)^j
\cot^{1/2} \left(\frac{(2j+1)\pi}{2(k+2)} \right) \ket{C,j},
\end{equation}
and the one located at $\pm \one$ is
\begin{equation}
\ket{C_\pm} = \left(\frac{2}{k+2} \right)^{\frac{1}{4}}
\sum_{j=0}^{k/2} 
\tan^{1/2} \left(\frac{(2j+1)\pi}{2(k+2)} \right) \ket{C,j}.
\end{equation}

\subsection{Geometry and M\"obius strips}

So far, the discussion of M\"obius strips has been restricted
to the evaluation and interpretation of amplitudes with the
boundary condition $\ket{0}$. In the $SU(2)$ case  we are
now going to interpret the  M\"obius strips with all boundary
conditions. This discussion is somewhat similar to that
in \cite{MSS} for $\BZ_2$ orbifolds of $SU(2)$.

Recall some basic facts about the geometry of the conjugacy
classes  \cite{ARSI, ARSII}. There are two conjugacy classes, $\pm\one$,
which are points, and $k-1$ conjugacy classes which are
spheres $S^2 \subset S^3$. Since we already evaluated the
M\"obius strips for the point-like branes we turn to the
conjugacy classes which are spheres. According to the analysis
of \cite{ARSI, ARSII} these spheres are ``fuzzy'' which amounts 
on the level of the
algebra of functions  to a truncation.
More precisely, the algebra of functions of a sphere is
spanned by spherical harmonics $Y^{j,m}$, where $j, m$ are
integers and $|m| \leq j$. For the fuzzy sphere
describing the geometry of the conjugacy class represented
by the group element $g_J$ on (\ref{gj}) the
label $j$ takes integer values up to a maximum value of $2J$.

In the case that the conjugacy class $J$ is either point-wise
or set-wise fixed under an involution, the fuzzy sphere algebra
inherits an involution from that reflection.
In our situation, the fuzzy sphere located at the equator is
the only sphere which is fixed under both involutions. More
precisely, it is point-wise fixed under $g \to -g^{-1}$ and
set-wise fixed under $g \to g^{-1}$. Under the latter involution,
individual group elements of the conjugacy class
get mapped as $g \to -g$. 
The inherited action on the algebra of spherical harmonics is
\begin{equation}\label{induced}
\begin{split}
Y^{jm} \to Y^{jm} \quad & {\rm for} \quad g \to -g^{-1}\\
Y^{jm} \to (-1)^j \ Y^{jm} \quad & {\rm for} \quad g \to g^{-1} .
\end{split}
\end{equation}

Turning to the conformal field theory point of view, the action
of the inherited involution on the spherical harmonics should
be compared to the action of $\Omega$ on the open string
sector primaries with appropriate boundary conditions. This
action can be read off from  the open string channel of the
M\"obius strip with boundary conditions $\ket{k/4}$, where
$\ket{k/4}$
is the Cardy state representing
the D-brane wrapping the equator:
\begin{equation}
\ket{\frac{k}{4}}= \left( \frac{2}{k+2} \right)^{\frac{1}{4}} \,
\sum_{j=0}^{k/2} (-1)^j \left( \sin \frac{\pi(2j+1)}{k+2}
\right)^{-\frac{1}{2}} \, \kket{j}.
\end{equation}
Note that this Cardy state exists only for $k$ even.
The cylinder amplitude for this boundary state contains all integer
representation labels,
\begin{equation}
Z_{cyl}^{k/4} = \sum_{j=0}^{k/2} \ \chi_j(\frac{it}{2}).
\end{equation}
According to the above geometric consideration, the M\"obius
strip should also contain all integer representation labels,
but this time with an insertion of $\Omega$, as determined
by equations (\ref{induced}). Indeed one can confirm that
\begin{equation}
M^{L}_{k/4} = \sum_{j=0}^{k/2} \ \chi_j(\frac{1+it}{2})
\longleftrightarrow \langle C_{eq}\ket{\frac{k}{4}},
\end{equation}
showing that the action of $\Omega $ on open string fields with
boundary condition $\ket{k/4}$
for the orientifold $\ket{C_{eq}}$ is trivial.

Similarly, the M\"obius strip for the crosscap $\ket{C_\pm}$ is:
\begin{equation}
M_{k/4}= (-1)^{\frac{k}{2}}
\sum_{j=0}^{k/2} (-1)^j \chi_j \longleftrightarrow 
\langle C_\pm \ket{\frac{k}{4}},
\end{equation}
showing that the projection (\ref{induced}) represents the
action of $\Omega$ on open string fields. More precisely,
(\ref{induced}) is induced by the crosscap $\ket{C_\pm}$ in the case
that $k=0$ mod $4$ and by $-\ket{C_\pm}$ in the case that $k=2$ mod $4$.

In the $SU(2)$ case, the involution $g\to g^{-1}$ actually leaves 
all conjugacy classes
set-wise fixed. One easily sees that the action on the representatives
$g_J$ written down in (\ref{gj}) is conjugation by the element
\begin{equation}
k= \left( \begin{array}{cc} 0&1 \\
                            -1&0
          \end{array}
   \right)
\end{equation}
All other elements in the conjugacy class $hg_Jh^{-1}$ can then be inverted
by conjugation with $hkh^{-1}$. In this way, the inversion $g\to g^{-1}$
can be understood as an inversion within the conjugacy classes. 
This is precisely what the M\"obius strip reflects: 
\begin{equation}
\sum_{j=0}^{2J} (-1)^j \chi_j \longleftrightarrow (-1)^{2J} \
\langle C_\pm \ket{J} = (-1)^{2(\frac{k}{2}-J)} \ \langle C_\pm \ket{\frac{k}{2} - J},
\end{equation}
where $J < k/4$. 
The representations
appearing in the amplitude are the same ones as for the cylinder 
amplitude with boundary conditions $J$ or $k/2 -J$. The reflection
is encoded in the non-trivial operation of $\Omega$ on those characters.
Note that the results for the M\"obius strip for the brane
labeled $\ket{J}$ and $\ket{k/2-J}$ agree. 
This is  easily understood in geometrical terms since
these pairs of branes have symmetric locations on the upper and
lower hemisphere of $S^2$. In particular, 
their location is symmetric with respect to the two
orientifold fixed point sets at the poles.
\medskip

For the other involution $g \to -g^{-1}$, all conjugacy classes
except the one located on the equator, get mapped to other conjugacy
classes. On the level of representatives, we
see that
\begin{equation}
g_J \to g_{k/2-J}.
\end{equation}.
Branes are mapped to image branes, and the M\"obius
amplitudes contain states propagating between a brane and its
image brane.
In the language of Cardy states, the image branes are obtained by
fusion of the Cardy label with the simple current,
mapping $\ket{J}$ to $\ket{k/2 -J}$. 

There are two  predictions for the M\"obius amplitudes arising
from these observations. The first one is that the M\"obius
amplitude with boundary condition $\ket{J}$ agrees with the
one with the image boundary condition $\ket{k/2 -J}$. 
The second one is that the characters appearing in the loop
channel of the M\"obius strip with boundary condition $\ket{J}$
should be the ones appearing in the cylinder with boundary conditions
$\ket{J}$ and $\ket{k/2 -J}$.

Indeed, an 
explicit computation of the M\"obius amplitudes yields the following 
result:
\begin{equation}
M^{J} = M^{\frac{k}{2}-J} = \sum_{l=0}^{k/2} \frac{1}{\sqrt{k+2}}  \
(-1)^l \ 
\frac{\sin\left(\pi\frac{(2l+1)(2J+1)}{k+2}\right)}
{\sin\left(\pi\frac{(2l+1)}{2(k+2)}\right)} \ \  \chi_l
\longleftrightarrow
\sum_{j=\frac{k}{2}-2J}^{k/2} \chi_j,
\end{equation}
confirming the geometrically motivated predictions.

This concludes the interpretation of all M\"obius strips in the
$SU(2)_k$ model in terms of geometric data.

\section{The shape computation}
The shape of the two orientifold planes can be determined
by scattering massless closed string states. 
Geometrical results are only expected in the limit $k\to \infty$.
This has been done for boundary states in \cite{MMSI, MMSII, FFFS}.
Since we will closely follow the discussion in \cite{MMSI, MMSII},
let us briefly review their argument.
Scattering amplitude between boundary- or crosscap states and
massless closed strings are computed as overlaps of the
boundary (crosscap) states with closed string ground states,
which are in the case of $SU(2)$ the states $\ket{j,m,m'}$,
where no descendant appears.
To determine the location of the orientifold,
it is useful to  pick a graviton wave packet localized at a point
of the group manifold.
To write down such a $\delta$-function shaped closed string
state, \cite{MMSI, MMSII} made use of the Peter-Weyl theorem which
states that the space of functions $\CF(G)$ on a group
manifold is isomorphic to an infinite direct sum of tensor products
of irreducible representations. The matrix elements of
the irreducible representations form a complete orthogonal
basis for the space of functions $\CF(G)$. For $SU(2)$ the
following rescaled basis is an orthonormal 
basis with respect to the Haar measure on the group:
\begin{equation}
\sqrt{2j+1} \   D^j_{m m'}(g) = \langle jm| R(g)\ket{jm'}
\end{equation}
In this basis a closed string $\delta$-function on the group manifold reads:
\begin{equation}\label{delta}
\ket{g} = \sum_{j} \sqrt{2j+1} \ket{j,m,m'}.
\end{equation}
However, one also wants that the closed string probe contains
only states with low $j$, $j^2 \leq k$, since only in that
case the closed string states are well localized. 
To suppress higher $j$ one can introduce an explicit cutoff
factor $\exp(-j^2/k)$ into the above sum, which suppresses
string modes with $j \geq \sqrt{k}$. 

In \cite{MMSI} the $SU(2)$ group manifolds is parametrized
by three angles,
summarized in a vector $\vec \theta$. The only angle which we
need explicitly is the angle $\psi$, which labels the $SU(2)$
conjugacy classes. If we want to relate this to our earlier
discussion at finite $k$, we can set $\psi = 2 \pi i J/k$, in particular,
$\psi= 0, \dots, \pi$, and $\one $ and $-\one$ are located at
$\psi= 0$ and $\psi=\pi$.
The other two angle variables parametrize the spheres $S^2 \subset S^3$,
and will not enter the discussion explicitly. A group element
in the conjugacy class labeled $\psi$ will be denoted $g_\psi$,
in analogy to our earlier notation $g_J$.
\cite{MMSI, MMSII} then proceed and take overlaps of boundary states
with $\delta$-function states (\ref{delta}). 
To evaluate the expressions, one needs
the relation of the matrix elements $D^j_{mm'}$ to the
$SU(2)$ characters:
\begin{equation}
\sum_m D_{mm}^j (g_\psi) = \frac{\sin(2j+1) \psi}{\sin \psi}.
\end{equation}

We now want to do an analogous computation for the crosscap
states $\ket{C_{eq}}$ and $\ket{C_\pm}$.
The difference
between the boundary state and the crosscap
computation is that  the boundary state is a linear combinations
of conventional Ishibashi states, whereas the crosscap is a
combination of the crosscap Ishibshi states (\ref{cblock}). However, since
the closed string states with which we take the overlap are
ground states, this does not modify the computation. 

Another important point to keep in mind is 
that the coefficients of the crosscap Ishibashi states can
only be non-vanishing when the primary on which the Ishibashi
state is built is invariant under $\Omega$. Therefore, the summation
is only over a subset of representations and the crosscap is less
well localized compared to  boundary states. 

In the case at hand only primaries with even $j$ enter the crosscap
state. $\Omega$-invariant $\delta$ functions (for which the summation in
(\ref{delta}) is only over even $j$) can resolve conjugacy classes
only up to reflections around the equator, meaning,
they cannot distinguish the location at $g_J$ from the location
at $g_{k/2-J}$. 
The consequence is of course that 
the orientifold loci have to be symmetric around the
equator. D-brane boundary states can only be localized up
to reflections around the equator by invariant $\delta$-functions.
This is another point of view one can take when
interpreting the M\"obius amplitudes computed earlier.

Keeping all this in mind, we can perform the analogous computation for 
crosscaps:
\begin{equation}
\begin{split}
\langle C_{eq} \ket{g_\psi} & \sim
\sum_{j=0}^{k/2} \sum_m (-1)^j \ \cot^{1/2}
\left( \pi \frac{2j+1}{2(k+2)} \right) \sqrt{2j+1} \ \ D^j_{mm'}(g_\psi)^* \\
& \sim \sum_j (-1)^j \frac{\sin(2j+1) \psi}{\sin \psi} \\
& \sim \delta(\psi - \frac{\pi}{2} ).
\end{split}
\end{equation}
The approximation made passing from the first to the second line
is only good if $j^2 << k$. 
One can make
this more explicit in the equations by using the
explicit cutoff $exp(-j^2/k)$. 
In this way we obtain the
``shape-function'' $s(\psi)$:
\begin{equation}
s(\psi) =  \left(\frac{2}{k+2} \right)^{\frac{1}{4}}
\sum_{j=0}^{k/2} e^{\frac{-j^2}{k}} (-1)^j \
\cot^{1/2} \left(\frac{(2j+1)\pi}{2(k+2)} \right) \sqrt{2j+1} \ \
\frac{\sin(2j+1)\psi}{\sin\psi}.
\end{equation}
This function has been plotted for $k=20$ in the following picture:
\vspace{1cm}
\fig{
The shape function $s(\psi)$ for $\ket{C_{eq}}$. $\psi$ is plotted in multiples of
$\pi$.}
{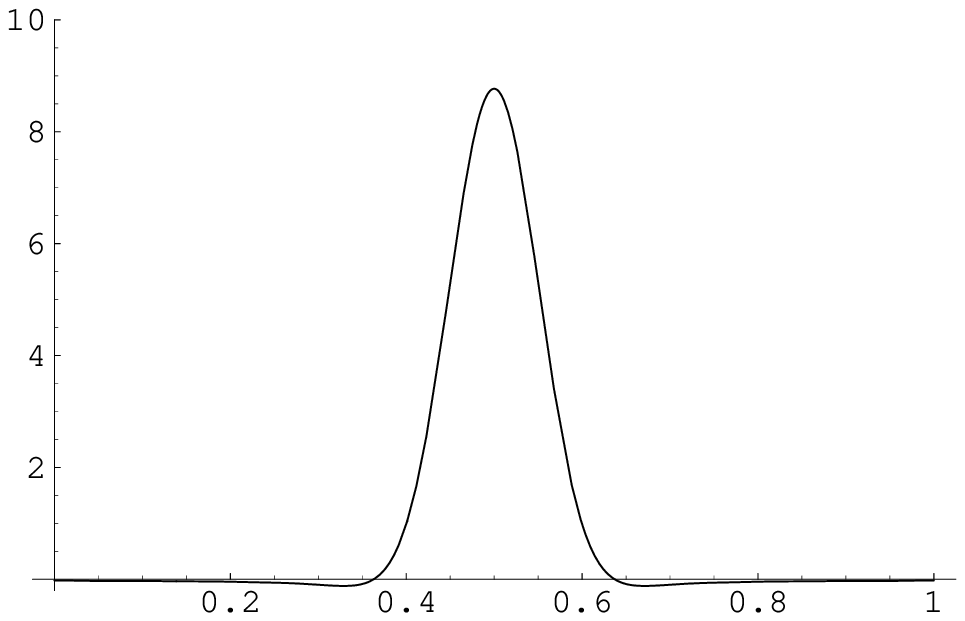}{8truecm}
\figlabel{\ddns}
\vspace{1cm}
This clearly shows that the orientifold locus is at $\psi = \pi/2$.
\medskip

Let us move on to the second crosscap. 
We first make the same approximations as before:
\begin{equation}
\begin{split}
\langle C_\pm \ket{g_\psi} & \sim
\sum_{j=0}^{k/2} \sum_m \tan^{1/2}
\left( \pi \frac{2j+1}{2(k+2)} \right) \sqrt{2j+1} \ \ D^j_{mm'}(g_\psi)^* \\
& \sim \sum_j  \sin \left( \frac{\pi(2j+1)}{k+2}\right) \ \ 
\frac{\sin(2j+1) \psi}{\sin \psi} \\
& \sim \delta(\psi - \pi) + \delta (\psi).
\end{split}
\end{equation}
Again, the approximation leading to the second line is  bad
unless big $j$ are explicitly suppressed.
The shape function for this case is given by:
\begin{equation}
s(\psi) = \left(\frac{2}{k+2} \right)^{\frac{1}{4}}
\sum_{j=0}^{k/2} e^{\frac{-j^2}{k}} 
\tan^{1/2} \left(\frac{(2j+1)\pi}{2(k+2)} \right) \sqrt{2j+1}
\frac{\sin(2j+1)\psi}{\sin\psi}.
\end{equation}
It is plotted in the following picture:
\vspace{1cm}
\fig{
The shape function for $\ket{C_\pm}$.}
{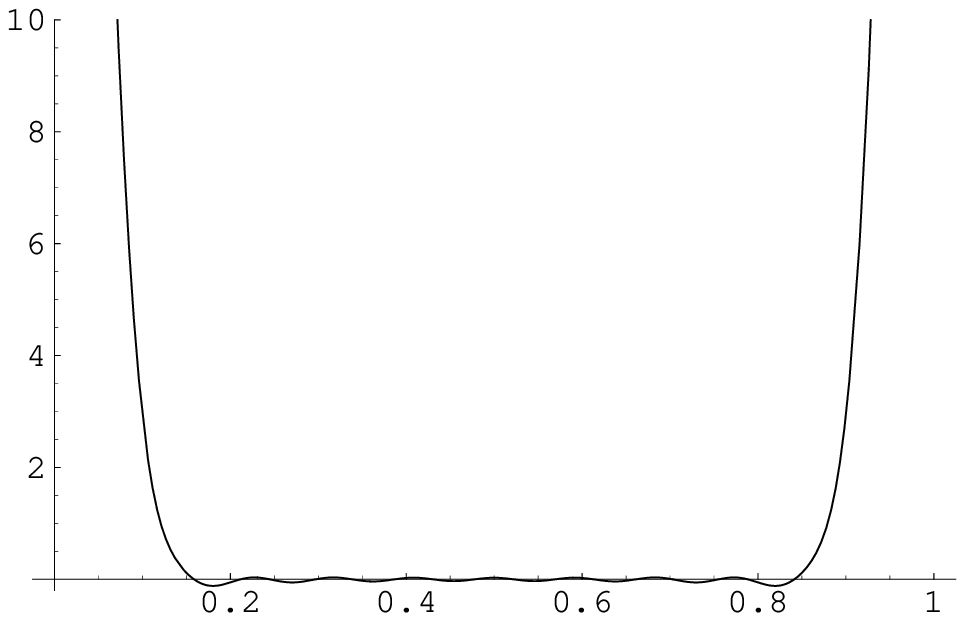}{8truecm}
\figlabel{\ddns}
\vspace{1cm}

\section*{Acknowledgements}
I would like to thank G.~Moore for initial collaboration
and subsequent discussions, in particular for suggesting
the computation in section 4.

\newpage
\bibliography{wzw}
\bibliographystyle{utphys}

\end{document}